\documentclass
[preprint,pre,10pt,twocolumn,superscriptaddress,notitlepage,showpacs,nofootinbib]{revtex4}%
\usepackage{amsfonts}
\usepackage{amsmath}
\usepackage{amssymb}
\usepackage{graphicx}%
\begin{document}
\title{Metropolis Monte Carlo algorithm based on reparametrization invariance}
\author{L. Velazquez}
\affiliation{Departamento de F\'{\i}sica, Universidad de Pinar del Rio, Marti 270, Esq. 27
de Noviembre, Pinar del Rio, Cuba}
\author{J.C. Castro Palacio}
\affiliation{Departamento de F\'{\i}sica, Universidad de Pinar del Rio, Marti 270, Esq. 27
de Noviembre, Pinar del Rio, Cuba}

\begin{abstract}
We introduce a modification of the well-known Metropolis importance sampling
algorithm by using a methodology inspired on the consideration of the
\textit{reparametrization invariance} of the microcanonical ensemble. The most
important feature of the present proposal is the possibility of performing a
suitable description of microcanonical thermodynamic states during the
first-order phase transitions by using this local Monte Carlo algorithm.

\end{abstract}
\date{\today}
\maketitle

\section{Introduction}

The basic problem in equilibrium statistical mechanics is to compute the phase
space average, in which Monte Carlo method plays a very important role
\cite{mc1,mc2,mc3}. Among all admissible statistical ensembles used with the
above purpose, the microcanonical ensemble provides the most complete
characterization of a given system in thermodynamic equilibrium. While
microcanonical calculations based on microscopic dynamics can suffer from the
existence of metastable states with large relaxation times \cite{lat1,lat2},
the available Monte Carlo methods based on reweighting technics (such as the
Multicanonical Monte Carlo \cite{berg1,berg2}) are also so much machine-time
consuming due to they try to obtain the density of states for all values of
energy in only one run. It is very well known that the microcanonical
observables can be suitably obtained by using the canonical ensemble whenever
\textit{ensemble equivalence} takes place in a sufficient large system.

One of the most famous and widely used Monte Carlo algorithm considering the
Gibbs canonical ensemble is the Metropolis importance sampling algorithm
\cite{metro}. This method has some important features. It is extremely general
and each moves involves $O\left(  1\right)  $ operations. However, its
dynamics suffers from \textit{critical slowing down,} that is, if $N$ is the
system size, the correlation time $\tau$ diverges as a critical temperature is
approached: the divergence follows a power law behavior $\tau\propto N^{z}$ in
second-order phase transitions, while $\tau$ diverges exponentially with $N$
in first-order phase transitions as consequence of the ensemble inequivalence
(multimodal character of the energy distribution function).

The aim of the present Letter is to develop a new methodology in which the
local algorithm Metropolis Monte Carlo could be used for performing
microcanonical calculations by avoiding the anomalous behaviors associated
with the presence of a first order phase transition in a given energetic
region. The key of our method is to consider certain generalized
canonical-like ensemble which is equivalent to the microcanonical ensemble for
$N$ large enough in spite of the presence of a first-order phase transition, a
methodology inspired by our recent paper \cite{vel}.

\section{The method}

Let us consider a given Hamiltonian system $\hat{H}$ becoming extensive in the
thermodynamic limit $N\rightarrow\infty$. As already commented, the basic idea
of our method is to substitute the Gibbs canonical ensemble by the following
\textit{generalized canonical ensemble}:%

\begin{equation}
\hat{\omega}_{c}\left(  \eta,N\right)  =\frac{1}{Z_{c}\left(  \eta,N\right)
}\exp\left(  -\eta\Theta\left(  \hat{H}\right)  \right)  , \label{gce}%
\end{equation}
where $\Theta\left(  E\right)  \equiv N\varphi\left(  E/N\right)  $, being
$\varphi\left(  \varepsilon\right)  $ certain analytical \textit{bijective
function} of the energy per particle of the system $\varepsilon=E/N$.

The \textit{Planck potential} $\mathcal{P}\left(  \eta,N\right)  =-\ln
Z_{c}\left(  \eta,N\right)  $ associated with the partition function:%
\begin{equation}
Z_{c}\left(  \eta,N\right)  =\int\exp\left(  -\eta N\varphi\left(  \frac{1}%
{N}E\right)  \right)  \Omega\left(  E,N\right)  dE, \label{partition}%
\end{equation}
can be estimated for $N$ large by using the steepest descend method as
follows:%
\begin{equation}
\mathcal{P}\left(  \eta,N\right)  \simeq N\left\{  \eta\varphi\left(
\varepsilon_{e}\right)  -s\left(  \varepsilon_{e}\right)  \right\}  -\frac
{1}{2}\ln\left(  \frac{2\pi\sigma_{\varepsilon}^{2}\left(  \varepsilon
_{e}\right)  }{N}\right)  , \label{Legendre}%
\end{equation}
where $s\left(  \varepsilon\right)  =\ln\left\langle W\left(  E,N\right)
\right\rangle /N$ is the entropy per particle of the system, being
$W=\Omega\left(  E,N\right)  \delta\varepsilon_{0}$ the microcanonical
accessible volume, and $\varepsilon_{e}$, the \textit{only one} stationary
energy of the maximization problem $\max_{\varepsilon_{e}}\left\{  \eta
\varphi\left(  \varepsilon\right)  -s\left(  \varepsilon\right)  \right\}  $
whose stationary conditions demand:%
\begin{align}
\eta\frac{\partial\varphi\left(  \varepsilon_{e}\right)  }{\partial
\varepsilon}  &  =\frac{\partial s\left(  \varepsilon_{e}\right)  }%
{\partial\varepsilon}\text{ and}\label{c1}\\
\frac{1}{\sigma_{\varepsilon}^{2}\left(  \varepsilon_{e}\right)  }  &
=\eta\frac{\partial^{2}\varphi\left(  \varepsilon_{e}\right)  }{\partial
\varepsilon^{2}}-\frac{\partial^{2}s\left(  \varepsilon_{e}\right)  }%
{\partial\varepsilon^{2}}>0. \label{c2}%
\end{align}
We found by combining (\ref{c1}) and (\ref{c2}) that the necessary and
sufficient condition for the existence of a unique stationary point of the
above maximization problem for any $\eta$ is given by:%
\begin{equation}
-\frac{\partial\varphi\left(  \varepsilon\right)  }{\partial\varepsilon}%
\frac{\partial}{\partial\varepsilon}\left\{  \left(  \frac{\partial
\varphi\left(  \varepsilon\right)  }{\partial\varepsilon}\right)  ^{-1}%
\frac{\partial s\left(  \varepsilon\right)  }{\partial\varepsilon}\right\}
>0, \label{c3}%
\end{equation}
which has to be applicable for all those admissible values of the energy per
particle $\varepsilon$. This condition can be rephrased by considering the
inverse function $\varepsilon\left(  \varphi\right)  $ of the bijective
function $\varphi\left(  \varepsilon\right)  $, and taking the entropy $s$ as
a \textit{scalar function} which is now rewritten in terms of the variable
$\varphi$:%
\begin{equation}
s\left(  \varphi\right)  \equiv s\left[  \varepsilon\left(  \varphi\right)
\right]  \Rightarrow-\left(  \frac{\partial\varphi\left(  \varepsilon\right)
}{\partial\varepsilon}\right)  ^{2}\frac{\partial^{2}s\left(  \varphi\right)
}{\partial\varphi^{2}}>0, \label{maincond}%
\end{equation}
which clarifies us that the entropy must be a concave function in this last parametrization.

The approximation (\ref{Legendre}) in the thermodynamic limit is just the
Legendre transformation between the thermodynamic potentials, $\mathcal{P}%
\left(  \eta,N\right)  \simeq\eta\Theta-S\left(  \Theta,N\right)  $ with
$S\left(  \Theta,N\right)  =S\left(  E,N\right)  $, which takes place as a
consequence of the equivalence between the generalized canonical ensemble
(\ref{gce}) and the microcanonical ensemble. \ Thus, \textit{we can ensure the
ensemble equivalence for all energy values by using an appropriate bijective
function }$\varphi\left(  \varepsilon\right)  $\textit{ without mattering
about the presence of a first-order phase transition}. This remark is
precisely the key of the Metropolis algorithm described below.

Considering the analogy with the Gibbs canonical ensemble, the generalized
canonical ensemble represents a system in which the parameter $\eta$
associated with the macroscopic observable $\Theta$ is keeping fixed by means
of certain external experimental arrangement. The state above is just a
consequence of the reparametrization invariance of the microcanonical
ensemble: the microcanonical description does not depend on the
reparametrization $\Theta=N\varphi\left(  E/N\right)  $ used for describing
the equilibrium thermodynamical states of the system. We only discuss in the
present work the applicability of this symmetry in improving the ordinary
Metropolis importance sampling algorithm based on the Gibbs canonical ensemble
(hereafter MMC), but the interested reader can see the recent paper \cite{vel}
about the implication of the reparametrization invariance in the searching of
a topological classification scheme of the phase transitions.

Generally speaking, the main difference between the MMC and the Metropolis
algorithm based on the reparametrization invariance (hereafter GCMMC) is the
using of the canonical-like weight $\omega\left(  E\right)  =\exp\left(
-\eta\Theta\left(  E\right)  \right)  $. Therefore, the probability $p$ for
the acceptance of a Metropolis move is given by:%
\begin{align}
p  &  =\min\left\{  1,\exp\left\{  -\eta\left[  \Theta\left(  E+\Delta
\varepsilon\right)  -\Theta\left(  E\right)  \right]  \right\}  \right\}
\nonumber\\
&  \simeq\min\left\{  1,\exp\left[  -\eta\frac{\partial\varphi\left(
\varepsilon\right)  }{\partial\varepsilon}\Delta\varepsilon-\frac{1}{2N}%
\eta\frac{\partial^{2}\varphi\left(  \varepsilon\right)  }{\partial
\varepsilon^{2}}\Delta\varepsilon^{2}\right]  \right\}  , \label{paso}%
\end{align}
in which all $O\left(  \frac{1}{N}\right)  $ contributions in the exponential
argument have been dismissed. While every energy decreasing $\Delta
\varepsilon<0$ is always accepted in the MMC, such probability decreases in
the GCMMC due to the presence of the quadratic term in the exponential
argument of (\ref{paso}), although such effect decreases with the $N$ increasing.

Obviously, when $N$ is large enough, the expression of the acceptance
probability of the GCMMC differs effectively from the one used in the MMC by
the consideration of a \textit{variable canonical parameter} $\beta\left(
\varepsilon;\eta\right)  $:%
\begin{equation}
\beta\left(  \varepsilon;\eta\right)  =\eta\frac{\partial\varphi\left(
\varepsilon\right)  }{\partial\varepsilon}, \label{can eff}%
\end{equation}
which fluctuates around the microcanonical value $\beta\left(  \varepsilon
_{e}\right)  =\partial s\left(  \varepsilon_{e}\right)  /\partial\varepsilon$
derived from the condition (\ref{c1}). Taking into account the sharp Gaussian
localization of the integral (\ref{partition}) in the thermodynamic limit, the
dispersion $\delta\beta\equiv\sqrt{\left\langle \delta\beta^{2}\right\rangle
}$ can be estimated in terms of the energy dispersion $\delta\varepsilon
\equiv\sqrt{\left\langle \delta\varepsilon^{2}\right\rangle }$ as follows:%
\begin{equation}
\delta\beta=\eta\frac{\partial^{2}\varphi\left(  \varepsilon_{e}\right)
}{\partial\varepsilon^{2}}\delta\varepsilon, \label{v1}%
\end{equation}
where $\left\langle \cdots\right\rangle $ denotes the average over the
ensemble (\ref{gce}). Considering the Gaussian estimation of the energy
dispersion as $\left\langle \delta\varepsilon^{2}\right\rangle =\sigma
_{\varepsilon}^{2}\left(  \varepsilon_{e}\right)  /N$ with $\sigma
_{\varepsilon}^{2}\left(  \varepsilon_{e}\right)  $ given in (\ref{c2}), the
equation (\ref{v1}) can be rewritten in the following form:%
\begin{equation}
\delta\beta=\frac{1}{N\delta\varepsilon}+\delta\varepsilon\frac{\partial
^{2}s\left(  \varepsilon\right)  }{\partial\varepsilon^{2}}. \label{v2}%
\end{equation}

This expression is a very nice result which talks about the limits of
precision in a general calculation of the caloric curve by using a local
algorithm based on the generalized canonical ensemble (\ref{gce}): while
temperature can be fixed ($\delta\beta=0$) wherever the entropy is a concave
function, the inverse temperature dispersion $\delta\beta$ can not vanish
wherever the entropy be a convex function in terms of energy $\varepsilon$.
Supercritical slowing down associated with the MMC method based on the Gibbs
canonical ensemble is precisely related with the downfall of the Gaussian
estimation of the energy dispersion $\delta\varepsilon=\left(  -N\partial
^{2}s\left(  \varepsilon\right)  /\partial\varepsilon^{2}\right)  ^{-\frac
{1}{2}}$ whenever the second derivative of the entropy goes to zero and
becomes nonnegative. It is very interesting to notice that the complementary
macroscopic observables energy and temperature obey the \textit{uncertainly
relation} $\delta E\delta\beta\geq1$ wherever $\partial^{2}S\left(
E,N\right)  /\partial E^{2}\geq0$, which is rather analogue to the Quantum
Mechanics uncertainly relation $\delta E\delta\tau\sim\hbar$. A better
analysis allows us to obtain the inequalities $\delta\varepsilon\leq\left(
-N\partial^{2}s\left(  \varepsilon\right)  /\partial\varepsilon^{2}\right)
^{-\frac{1}{2}}$ whenever entropy $s\left(  \varepsilon\right)  $ be a concave
function, while $\delta\beta\geq\left(  \partial^{2}s\left(  \varepsilon
\right)  /\partial\varepsilon^{2}/N\right)  ^{\frac{1}{2}}$ when entropy is convex.

As already commented, the success of the GCMMC methods relies on the selection
of a good bijective function $\varphi\left(  \varepsilon\right)  $ satisfying
the condition (\ref{maincond}), which obviously can be done by following
different schemes. For example, we can rephrase the stationary condition
(\ref{c1}) as $\eta=\partial s\left(  \varphi\right)  /\partial\varphi$ and
substitute it in (\ref{maincond}). The resulting equation expresses the
monotonic decreasing of the dependence $\eta$ \textit{versus} the energy per
particle $\varepsilon$:
\begin{equation}
\frac{\partial\eta\left(  \varepsilon\right)  }{\partial\varepsilon}<0,
\end{equation}
which talks about a bijective correspondence between $\eta$ and $\varepsilon$
as a consequence of the ensemble equivalence. This last demand suggests us the
implementation of an indirect Monte Carlo method based on the generalized
canonical ensemble (\ref{gce})\ where an arbitrary dependence $\eta\left(
\varepsilon\right)  $ satisfying the above condition be assumed \textit{a
priory}, but the unknown associated bijective function $\varphi\left(
\varepsilon\right)  $ must be reconstructed by using reweighting schemes
analogue to the ones used in the Multicanonical calculations
\cite{berg1,berg2}. Nevertheless, we are interested in the present work in the
implementation of a variant of the Metropolis algorithm with acceptance
probability (\ref{paso}) where the bijective function $\varphi\left(
\varepsilon\right)  $ will be proposed \textit{a priory} in order to avoids
the ensemble inequivalence in a Hamiltonian system with short-range interactions.

Let us assume that the interest system\ exhibits an ensemble inequivalence
inside the energy interval $\left(  \varepsilon_{1},\varepsilon_{2}\right)  $,
where $\partial^{2}s\left(  \varepsilon_{i}\right)  /\partial\varepsilon
^{2}<0$. The microcanonical parameter $\beta\left(  \varepsilon\right)
=\partial s\left(  \varepsilon\right)  /\partial\varepsilon$ will change
slowly with the energy $\varepsilon$ in this region due to $\left\vert
\partial^{2}s\left(  \varepsilon\right)  /\partial\varepsilon^{2}\right\vert
\simeq0$ in the first-order phase transitions (see in the application example
described in the next section), so that, $\eta\left(  \varepsilon\right)
\partial\varphi\left(  \varepsilon\right)  /\partial\varepsilon\simeq\beta
_{c}$ where $\beta_{c}$ is the inverse critical temperature. Assuming this
last relation as a first approximation, its substitution in the condition of
the ensemble equivalence (\ref{c2}) yields:
\begin{equation}
\frac{\partial\varphi\left(  \varepsilon\right)  }{\partial\varepsilon}%
=\frac{\beta_{c}}{\eta\left(  \varepsilon\right)  }\Rightarrow\frac{1}%
{\sigma_{\varepsilon}^{2}\left(  \varepsilon\right)  }\simeq\beta_{c}%
\lambda\left(  \varepsilon\right)  -\frac{\partial^{2}s\left(  \varepsilon
\right)  }{\partial\varepsilon^{2}}>0, \label{condition_lambda}%
\end{equation}
with $\lambda\left(  \varepsilon\right)  =-\partial\ln\eta\left(
\varepsilon\right)  /\partial\varepsilon$. This demands is very easy to
satisfy by considering the function $\lambda\left(  \varepsilon\right)  $ as a
large enough positive constant, $\lambda\left(  \varepsilon\right)
\equiv\lambda>0$. This assumption allows us to propose the first derivative of
the bijective function $\varphi\left(  \varepsilon\right)  $ as follows:%
\begin{equation}
\xi\left(  \varepsilon\right)  =\frac{\partial\varphi\left(  \varepsilon
\right)  }{\partial\varepsilon}=\left\{
\begin{array}
[c]{cc}%
1 & \text{if }\left\langle \varepsilon\right\rangle >\varepsilon_{2},\\
\exp\left(  -\lambda\left(  \varepsilon_{2}-\varepsilon\right)  \right)  &
\text{if }\left\langle \varepsilon\right\rangle \in\left(  \varepsilon
_{1},\varepsilon_{2}\right)  ,\\
\exp\left(  -\lambda\left(  \varepsilon_{2}-\varepsilon_{1}\right)  \right)  &
\text{otherwise,}%
\end{array}
\right.  \label{chita}%
\end{equation}
which should avoid the ensemble inequivalence within the energetic range
$\left(  \varepsilon_{1},\varepsilon_{2}\right)  $. Since $\xi\left(
\varepsilon\right)  $ is a constant when $\left\langle \varepsilon
\right\rangle >\varepsilon_{2}$ or $\left\langle \varepsilon\right\rangle
<\varepsilon_{1}$, the GCMMC becomes\ in the MMC algorithm outside the
interval $\left(  \varepsilon_{1},\varepsilon_{2}\right)  $.

Although the piecewise function $\xi\left(  \varepsilon\right)  $ is referred
in terms of the instantaneous values of the energy $\varepsilon$ in the
Metropolis dynamics, we propose in\ (\ref{chita}) that the selection of the
interval during the whole computation of the microcanonical averages at a
given energy depends on the average energy $\left\langle \varepsilon
\right\rangle $ instead of the instantaneous values $\varepsilon$. It can be
numerically checked that the existence of any abrupt change in the first
derivatives of the function $\xi\left(  \varepsilon\right)  $ in the
acceptance probability (\ref{paso}) can provoke a significant perturbation in
the convergence of the second derivative of the entropy (associated with the
heat capacity) by using the GCMMC when the instantaneous energy $\varepsilon$
is close to $\varepsilon_{1}$ or $\varepsilon_{2}$. However, the using of
$\left\langle \varepsilon\right\rangle $ instead of the instantaneous value
$\varepsilon$ allows the function $\xi\left(  \varepsilon\right)  $ to exhibit
a smooth behavior beyond the interest energy interval $\left(  \varepsilon
_{1},\varepsilon_{2}\right)  $ during the Metropolis dynamics.

In order to obtain a caloric curve $\beta$ \textit{versus} $\varepsilon$ with
approximately $M$ points with $\left\langle \varepsilon\right\rangle $ more or
less uniformly distributed in the interval $\left(  \varepsilon_{1}%
,\varepsilon_{2}\right)  $, the canonical parameter $\eta$ can be increased as
follows:%
\begin{equation}
\eta_{i+1}=\left\{
\begin{array}
[c]{cc}%
\eta_{i}+c & \text{if }\left\langle \varepsilon\right\rangle >\varepsilon
_{2},\\
\eta_{i}\exp\left(  \frac{1}{M}\lambda\left(  \varepsilon_{2}-\varepsilon
_{2}\right)  \right)  & \text{if }\left\langle \varepsilon\right\rangle
\in\left(  \varepsilon_{1},\varepsilon_{2}\right)  ,\\
\eta_{i}+c\exp\left(  \lambda\left(  \varepsilon_{2}-\varepsilon_{1}\right)
\right)  & \text{otherwise,}%
\end{array}
\right.  \label{eta}%
\end{equation}
where $c>0$ in the $\beta$ step outside the interest energy interval $\left(
\varepsilon_{1},\varepsilon_{2}\right)  $.

The reader may notice by reexamining the equation (\ref{condition_lambda})
that a very large value of the numeric constant $\lambda$ leads to a very
small energy dispersion $\delta\varepsilon$, provoking\ in this way an
increasing of the inverse temperature dispersion $\delta\beta$ as a
consequence of the uncertainly relation (\ref{v2}). An appropriate
prescription of $\lambda$ in order to minimize the dispersion $\left\langle
\delta p^{2}\right\rangle =\left\langle \delta\beta^{2}+\delta\varepsilon
^{2}\right\rangle $ in the caloric curve ($\varepsilon$ and $\beta$ in
dimensionless units) is given by:%
\begin{equation}
\beta_{c}\lambda\gtrapprox1. \label{w0}%
\end{equation}

Taking into consideration all the above exposed, microcanonical caloric curve
$\beta\left(  \varepsilon\right)  =\partial s\left(  \varepsilon\right)
/\partial\varepsilon$ can be obtained by using our GCMMC methods as follows:%
\begin{equation}
\beta\left[  \left\langle \varepsilon\right\rangle \pm\delta\varepsilon
\right]  =\eta\left\langle \xi\left(  \varepsilon\right)  \right\rangle
\pm\delta\beta, \label{w1}%
\end{equation}
where
\begin{align}
\delta\beta^{2}  &  =\frac{\sigma_{\beta}^{2}}{N}\equiv\eta^{2}\left(
\left\langle \xi^{2}\left(  \varepsilon\right)  \right\rangle -\left\langle
\xi\left(  \varepsilon\right)  \right\rangle ^{2}\right)  ,~\label{w2a}\\
\delta\varepsilon^{2}  &  =\frac{\sigma_{\varepsilon}^{2}}{N}\equiv
\left\langle \varepsilon^{2}\right\rangle -\left\langle \varepsilon
\right\rangle ^{2}, \label{w2b}%
\end{align}
while the second derivative of the entropy or curvature $\kappa\left(
\varepsilon\right)  =\partial^{2}s\left(  \varepsilon\right)  /\partial
\varepsilon^{2}$ can be given by:%
\begin{equation}
\kappa\left(  \left\langle \varepsilon\right\rangle \right)  \simeq\left(
\sigma_{\varepsilon}\sigma_{\beta}-1\right)  /\sigma_{\varepsilon}^{2},
\label{w3}%
\end{equation}
in accordance with the equation (\ref{v2}). The error $\delta\kappa$ in the
curvature after $n$ Metropolis iterations can be estimated by the formula:
\begin{equation}
\delta\kappa\simeq\left\langle 2\left\vert \kappa\right\vert +1/\sigma
_{\varepsilon}^{2}\right\rangle \sqrt{\frac{8\tau}{n}},
\end{equation}
where $\tau$ is the decorrelation time, that is, the minimum of Monte Carlo
iterations necessary to generate effectively independent, identically
distributed samples in the Metropolis dynamics.

\section{A simple application}

Let us apply the GCMMC method in order to perform the microcanonical
description of a Potts model \cite{pottsm}:%

\begin{equation}
H=\sum_{\left(  i,j\right)  }\left\{  1-\delta_{\sigma_{i}\sigma_{j}}\right\}
, \label{h}%
\end{equation}
on a two dimensional lattice (here with periodic boundary conditions) of
$N=L\times L$ spins with $q=10$ possible values (components). The sum is over
pairs of nearest neighbor lattice points only and $\sigma_{i}$ is the spin
state at the \textit{i-th} lattice point. Generally speaking, this model
system admits a ferromagnetic interpretation by introducing the bidimensional
vector variables $\mathbf{s}_{i}=\left[  \cos\left(  \kappa_{q}\sigma
_{i}\right)  ,\sin\left(  \kappa_{q}\sigma_{i}\right)  \right]  $ with
$\kappa_{q}=2\pi/q$, and defining the total magnetization\ as follows
$\mathbf{M}=\sum_{i}\mathbf{s}_{i}$.

As elsewhere shown, the $q=10$ states Potts model exhibits a first-order phase
transition associated to the ensemble inequivalence, which provokes the
existence of a supercritical slowing down during the ordinary Metropolis
dynamics based on the consideration of the Gibbs canonical ensemble. Clusters
algorithms, like Swendsen-Wang or Wolf algorithms \cite{wolf}, do not help in
this case \cite{gore} due to they are still based on the consideration of the
canonical ensemble, and consequently, the supercritical slowing down
associated to the ensemble inequivalence persists \cite{wang2}.%

\begin{figure}
[t]
\begin{center}
\includegraphics[
height=3.915in,
width=3.2396in
]%
{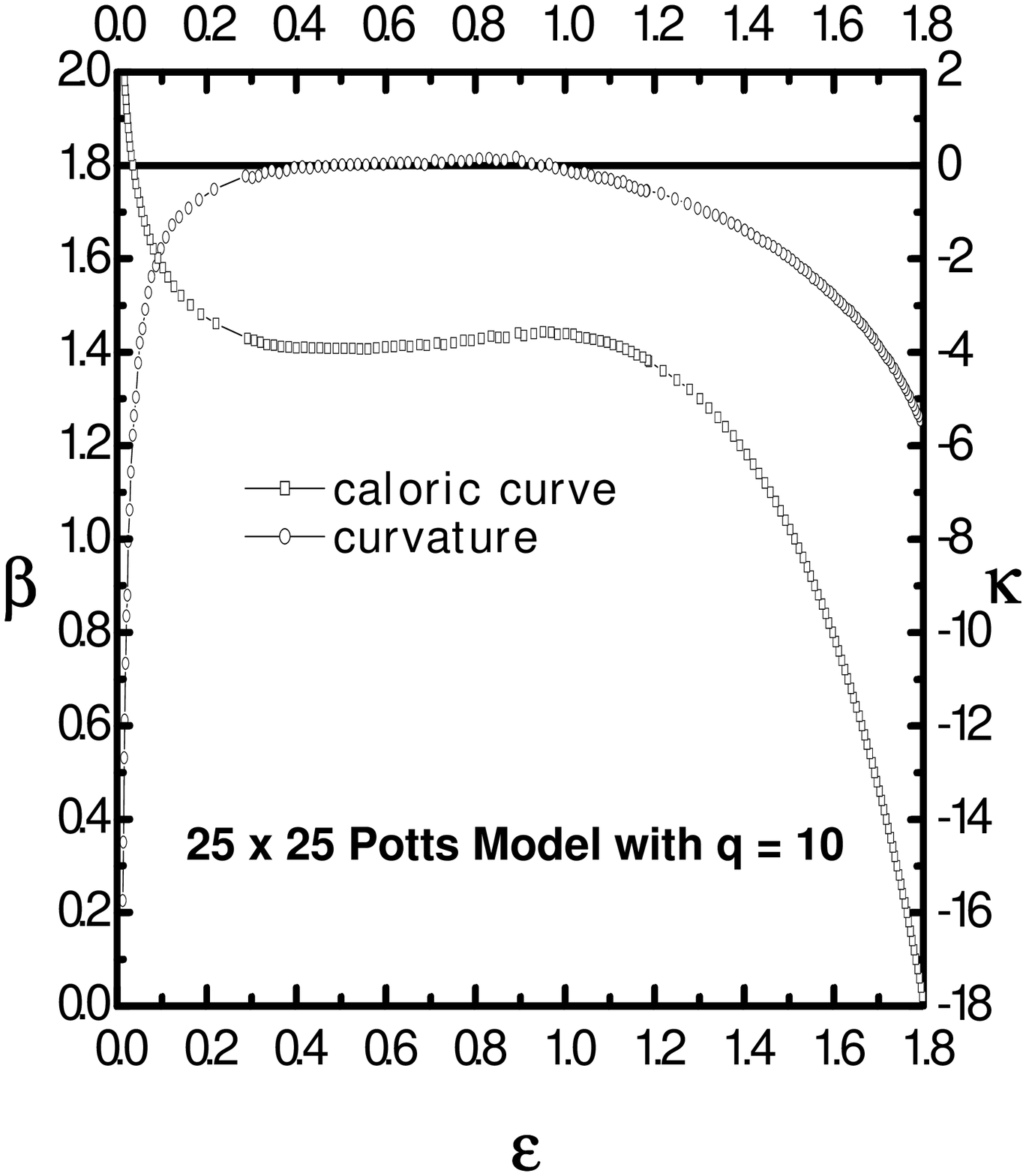}%
\caption{Microcanonical caloric curve $\beta\left(  \varepsilon\right)
=\partial s\left(  \varepsilon\right)  /\partial\varepsilon$ and curvature
$\kappa\left(  \varepsilon\right)  =\partial^{2}s\left(  \varepsilon\right)
/\partial\varepsilon^{2}$ of the $25\times25$ Potts model with $q=10$ states
with periodic boundary conditions obtained by using the GCMMC algorithm.
Notice the energetic region with a negative heat capacity. Errors are smaller
than the symbols linear dimension.}%
\label{potts}%
\end{center}
\end{figure}

This difficulty is successfully overcome by using the Multicanonical Monte
Carlo method \cite{berg1,berg2}, which reduces the exponential divergence of
the correlation times with respect to system size to a power at the
first-order phase transitions \cite{berg3}. Multicanonical ensemble flattens
out the energy distribution, which allows the computation of the density of
states $\Omega\left(  E,N\right)  $ for all values of $E$ in only one run.
This feature of the Multicanonical ensemble becomes a disadvantage when we are
also interested in the direct computation of the microcanonical average of
other microscopic observables at a given value of the energy, i.e. the
magnetization density dependence $\mathbf{m}\left(  \varepsilon\right)
=\left\langle \mathbf{M}\right\rangle /N$ of the Potts model (\ref{h}) with
$q=10$. This aim can be easily performed by using the GCMMC algorithm.%

\begin{figure}
[t]
\begin{center}
\includegraphics[
height=2.9611in,
width=3.5129in
]%
{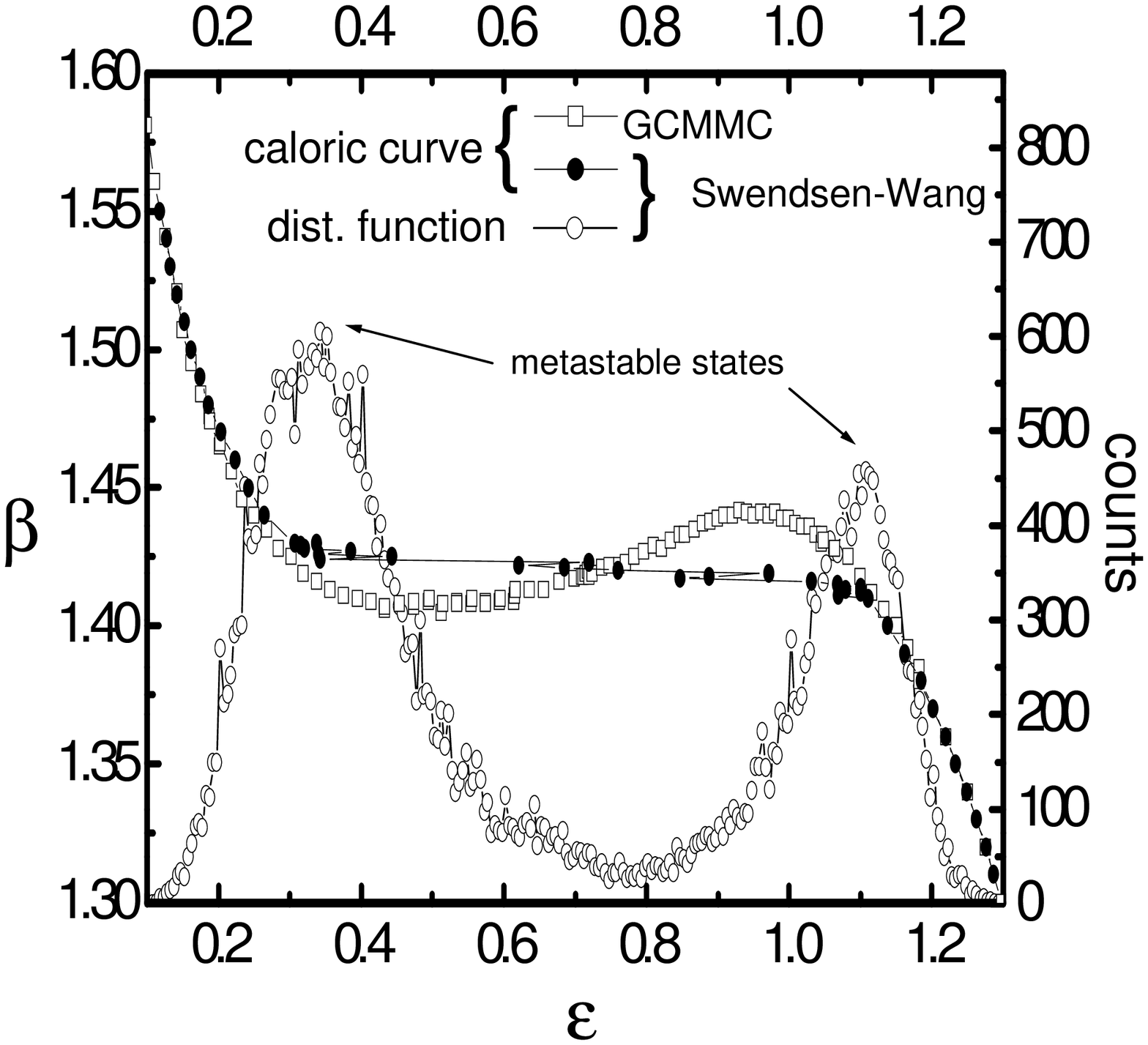}%
\caption{Comparative study between the GCMMC and the Swendsen-Wang (SW)
cluster algorithm. The SW algorithm is unable to describe the microcanonical
states with a negative heat capacity. Notice the bimodal character of the
energy distribution function at $\beta=1.42$. }%
\label{gc_sw.eps }%
\end{center}
\end{figure}

A preliminary calculation by using the MMC allows us to set the interest
energetic window with $\varepsilon_{1}=0.2$ and $\varepsilon_{2}=1.2$ in which
takes place the critical slowing down associated with the existence of a
first-order phase transition in this model for $L=25$. The inverse critical
temperature was estimated as $\beta_{c}\simeq1.4$, allowing us to set
$\lambda=0.8$. Besides, we also set $M=50$ and $c=0.02$ in the $\eta$
increasing given by the equation (\ref{eta}). The caloric $\beta\left(
\varepsilon\right)  $ and curvature $\kappa\left(  \varepsilon\right)  $
curves obtained by using the GCMMC algorithm with $n=10^{5}$ Metropolis
iterations for each point is shown in the FIG.\ref{potts}. A comparative study
between the present method and the Swendsen-Wang clusters algorithm
\cite{wang2} is shown in the FIG.\ref{gc_sw.eps }.

The backbending in the caloric curve $\beta$ \textit{versus} $\varepsilon$ is
directly associated to the existence of a negative heat capacity $c\left(
\varepsilon\right)  =-\beta^{2}\left(  \varepsilon\right)  /\kappa\left(
\varepsilon\right)  $ when $\varepsilon\in\left(  \varepsilon_{a}%
,\varepsilon_{b}\right)  $ with $\varepsilon_{a}=0.51$ and $\varepsilon
_{b}=0.93$, which is a feature of a first-order phase transition in a small
system becoming extensive in the thermodynamic limit. Since the heat capacity
is always positive within the canonical ensemble, $c_{c}=N\beta^{2}%
\left\langle \Delta\varepsilon^{2}\right\rangle _{c}\geq0$, such anomalous
regions are inaccessible in this description ($\left\langle \cdots
\right\rangle _{c}$ denotes the canonical average). This fact evidences the
existence of a significant \textit{lost of information }about the
thermodynamical features\ of the system during the occurrence of a first-order
phase transitions when the canonical ensemble is used instead of the
microcanonical one. This difficulty is successfully overcome by the GCMMC
algorithm, which is able to predict the microcanonical average of the
microscopic observables in these anomalous regions where any others Monte
Carlo methods based on the consideration of the Gibbs canonical ensemble such
as the original MMC, the Swendsen-Wang and the Wolff single cluster algorithms
certainly do not work.
\begin{figure}
[t]
\begin{center}
\includegraphics[
height=3.0113in,
width=3.2119in
]%
{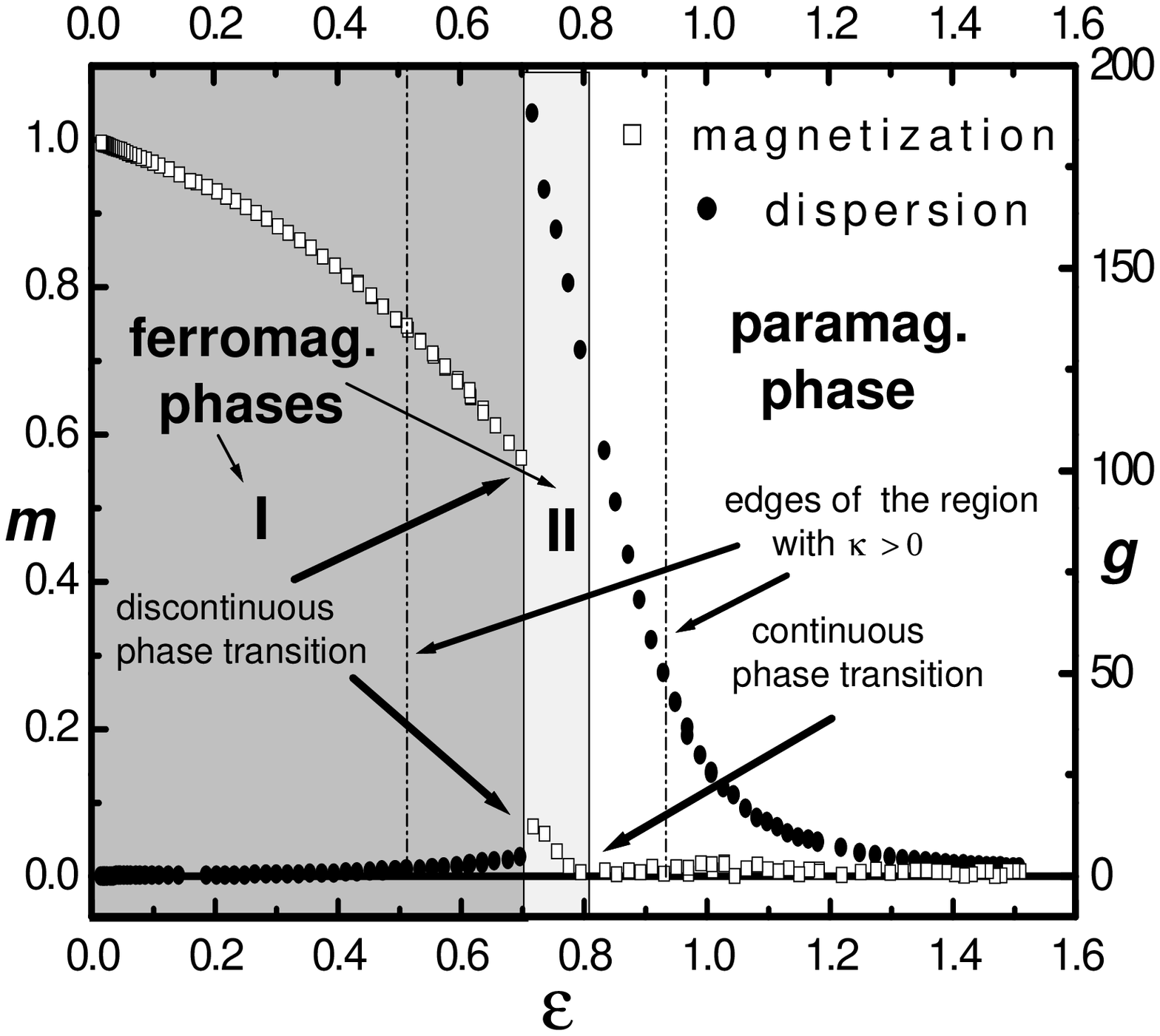}%
\caption{Magnetic properties of the model within the microcanonical ensemble
in which is shown the existence of two phase transitions at $\varepsilon
_{ff}\simeq0.7$ (ferro-ferro) and $\varepsilon_{fp}\simeq0.8$ (ferro-para).}%
\label{magnet1.eps}%
\end{center}
\end{figure}

This fact is clearly illustrated in the FIG.\ref{gc_sw.eps }, which evidences
that the Swendsen-Wang algorithm is unable to describe the thermodynamic
states with a negative heat capacity: its results within the anomalous region
differ significantly from the ones obtained by using the GCMMC algorithm. The
Swendsen-Wang dynamics exhibits here an erratic behavior originated from the
competition of the two metastable states present in the neighborhood of the
critical point (the energy distribution function in the canonical ensemble is
bimodal when $\beta\in\left(  \beta_{1},\beta_{2}\right)  $, where $\beta
_{1}=1.405$ and $\beta_{2}=1.445$, being this feature the origin of the
supercritical slowing down).%

\begin{figure}
[ptb]
\begin{center}
\includegraphics[
height=2.911in,
width=3.2119in
]%
{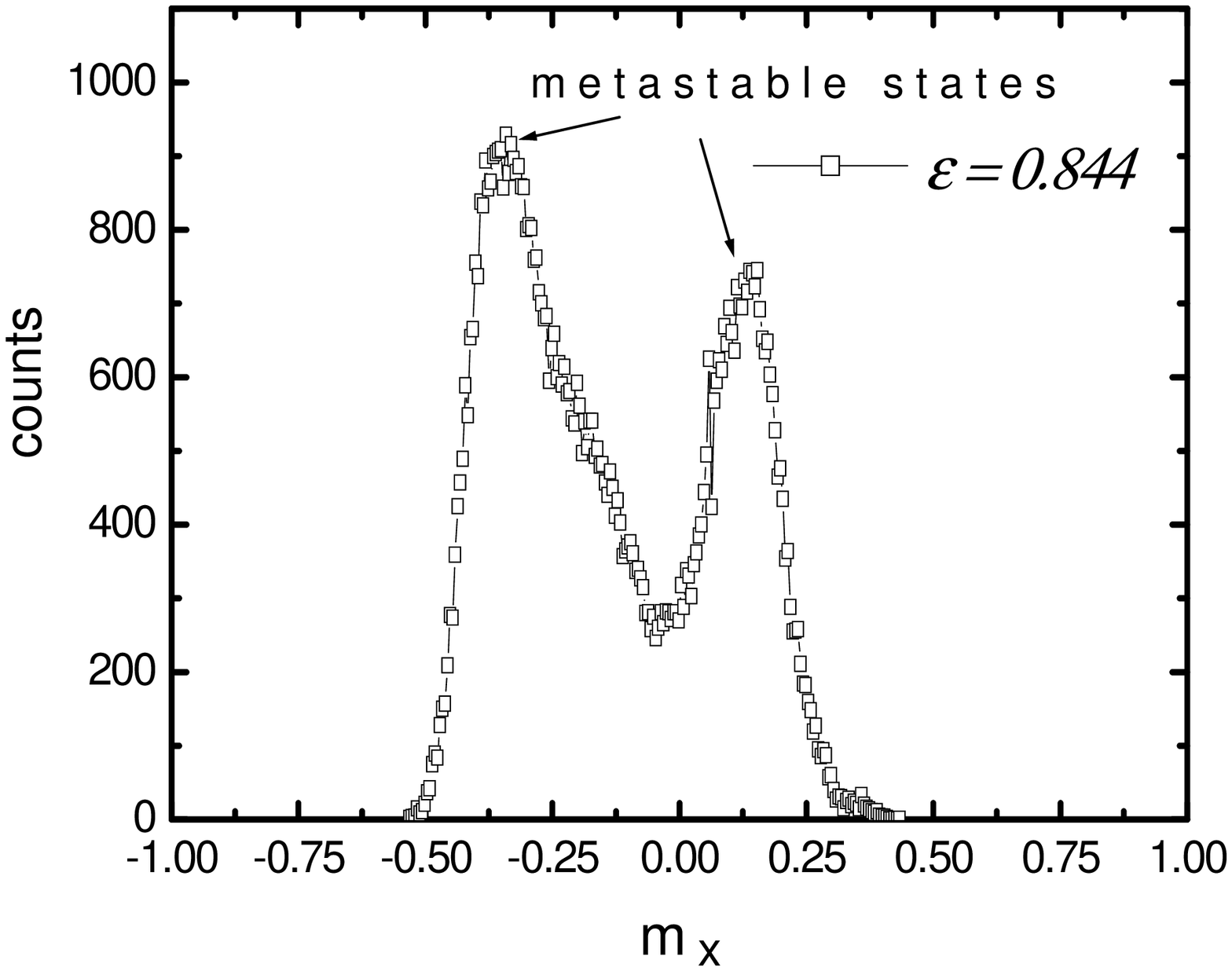}%
\caption{Histogram of the projection of the microscopic magnetization along
the direction of the spontaneous magnetization of the ferromanet type I, which
demostrates the existence of metastable states when $\varepsilon=0.844$.}%
\label{metastable.eps}%
\end{center}
\end{figure}

As already shown by Gross in ref.\cite{gro na}, a negative heat capacity in a
system with short-range interactions can be associated to the existence of a
non-vanishing interphase surface tension. Other quantities like the transition
temperature $\beta_{cr}$ and the latent heat $q_{lat}$ can be derived easily
from the caloric curve $\beta\left(  \varepsilon\right)  $. Following the same
procedure developed by Gross, our calculations allow us to obtain the
following estimates for the $q=10$ states Potts model with $L=25$: $\beta
_{cr}\simeq1.421$ and $q_{lat}\simeq0.78$.

The study of the magnetic properties of this model within the microcanonical
ensemble illustrated in the FIG.\ref{magnet1.eps}, shows us that the existence
of the non-vanishing interphase surface tension is not the only one
information hidden behind of a negative heat capacity within the canonical
description. The modulus of the magnetization density $m\left(  \varepsilon
\right)  $ evidences what could be considered as the signature of two
\textit{phase transitions within the microcanonical description of this model
system}: a continuous (paramagnetic-ferromagnetic) phase transition at the
critical point $\varepsilon_{fp}\simeq0.8$, and a discontinuous
(ferro-ferro)\ phase transition at $\varepsilon_{ff}\simeq0.7$ (from a low
magnetized ferromagnetic type II phase towards a high magnetized ferromagnetic
type I one). The investigation of these interesting behaviors deserves a
further study.

Most of thermodynamic points of this dependences were obtained from a data of
$n=10^{5}$ Metropolis iterations, with the exception of all those points
belonging to the energetic interval $\left(  0.7,0.93\right)  $ where very
large fluctuations of the magnetization density were observed. The possibility
of the GCMMC algorithm of performing a localized computation of the
microcanonical averages at a given value of the energy allows us to increase
the Metropolis iterations $n$ up to $5\times10^{6}$ for each point within the
above energetic range in order to reduce the significant dispersion of the
expectation values observed there.

The very large fluctuations, the peculiar behavior of the dispersion $g\left(
\varepsilon\right)  =\left\langle \left(  \mathbf{M-}\left\langle
\mathbf{M}\right\rangle \right)  ^{2}\right\rangle /N$, the qualitative form
of the magnetization curve $m\left(  \varepsilon\right)  $ shown in
FIG.\ref{magnet1.eps}, as well as the apparent large relaxation times of the
expectation values during the GCMMC dynamics, suggest us the presence of
several metastable states with different magnetization densities at a given
energy within this last region, whose existence is shown in the
FIG.\ref{metastable.eps}.

The anomalies observed in the thermodynamic characterization of the $q=10$
state Potts model suggest strongly the existence of the critical slowing down
phenomena within the "microcanonical description" provided by the GCMMC
algorithm. This fact inspires the development of nonlocal Monte Carlo
algorithms based on the consideration of the generalized canonical ensemble
(\ref{gce}) in order to address the microcanonical description of more larger
systems undergoing a discontinuous (first-order) phase transitions within the
canonical description.

\section{Conclusions}

We have presented a very simple methodology for improving the ordinary
Metropolis importance sampling algorithm \cite{metro} in order to allow the
computation of the microcanonical averages during the occurrence of the
first-order phase transitions in systems with short-range interactions. The
key of our approach is the consideration of certain generalized canonical
ensemble (\ref{gce})\ inspired on the reparametrization invariance of the
microcanonical description \cite{vel}, which becomes equivalent to the
microcanonical ensemble when the size $N$ of the interest system is large enough.

A direct advantage of this new methodology is the possibility of performing a
localized computation of the any microcanonical average at a given value of
the energy, and avoid in this way the unnecessary computation of the whole
energy range, a feature of many other Monte Carlo technics based on
reweighting the energy histogram \cite{BA1,lee,sw}. The existence of anomalies
within the microcanonical description of the $q=10$ states Poots model
inspires the further development of nonlocal Monte Carlo algorithms based on
the consideration of the generalized canonical ensemble (\ref{gce}).

\begin{acknowledgments}
L. Velazquez and J.C. Castro-Palacios thank the financial support of this
investigation project with code number 16-2004 obtained from the Cuban
\textbf{PNCB}.
\end{acknowledgments}


\begin{thebibliography}{99}                                                                                               %


\bibitem {mc1}M. H. Kalos and P. A. Whitlock, \textit{Monte Carlo Methods} Vol
I: Basics (John Wiley \& Sons , 1986).

\bibitem {mc2}G. S. Fishman, \textit{Monte Carlo, concepts, algorithms, and
applications} (Springer, 1996).

\bibitem {mc3}P. D. Landau and K. Binder, \textit{A guide to Monte Carlo
simulations in Statistical Physics} (Cambridge Univ Press, 2000).

\bibitem {lat1}V. Latora, A. Rapisarda and S. Ruffo, Phys. Rev. Lett.
\textbf{83} (1999) 2104; Physica A \textbf{280} (2000) 81; Physica D
\textbf{131} (1999) 38; Nucl. Phys. A \textbf{681} (2001) 331c.

\bibitem {lat2}V. Latora and A. Rapisarda, Prog. Theor. Phys. Suppl.
\textbf{139} (2000) 204.

\bibitem {berg1}B. A. Berg, J. Stat. Phys. \textbf{82} (1996) 323.

\bibitem {berg2}B. A. Berg, Fields Inst. Commun. \textbf{26} (2000) 1.

\bibitem {metro}N. Metropolis, A. W. Rosenbluth, M. N. Rosenbluth, A. H.
Teller and E. Teller, J. Chem. Phys. \textbf{21} (1953) 1087.

\bibitem {vel}L. Velazquez and F. Guzman, submitted to Phys. Rev. Lett.;
e-print (2006) [cond-mat/0604487] and references therein.

\bibitem {pottsm}J. -S. Wang, R. H. Swendsen and R. Koteck\'{y}, Phys. Rev.
Lett. \textbf{63} (1989) 1009.

\bibitem {wolf}U. Wolff, Phys. Rev. Lett. \textbf{62} (1989) 361.

\bibitem {gore}V. K. Gore and M. R. Jerrum, Proceeding of the 29th Anual ACM
Symposium on Theory of Computing (1997) 674; J. Stat. Phys. \ \textbf{97}
(1999) 67.

\bibitem {wang2}J. S. Wang, \textit{Efficient Monte Carlo Simulations Methods
in Statistical Physics}, e-print (2006) [cond-mat/0103318].

\bibitem {berg3}B. A. Berg and T. Neuhaus, Phys. Rev. Lett. \textbf{68} (1992) 9.

\bibitem {gro na}D.H.E Gross and M. E. Madjet, Z. Physic B \textbf{104} (1997)
521; e-print (1997) [cond-mat/9707100].

\bibitem {BA1}P. M. C. de Oliveira, Eur. Phys. J. B \textbf{6} (1998) 111.

\bibitem {lee}J. S. Wang and L. W. Lee, Comp. Phys. Commu. \textbf{127} (2000)
131; J. S. Wang, Physica A \textbf{281} (2000) 174.

\bibitem {sw}R. H. Swendsen, B. Diggs, J. S. Wang, S. T. Li, C. Genovese and
J. B. Kadane, Int. J. Mod. Phys. C \textbf{10} (1999) 1563.
\end{thebibliography}
\end{document}